\documentclass[fleqn,usenatbib]{mnras}

\usepackage{mathptmx}

\usepackage[T1]{fontenc}

\DeclareRobustCommand{\VAN}[3]{#2}
\let\VANthebibliography\thebibliography
\def\thebibliography{\DeclareRobustCommand{\VAN}[3]{##3}\VANthebibliography}

\def\ks{km s$^{-1}$}
\def\d{$^\circ$}
\def\m{$^\prime$}
\def\s{$^{\prime\prime}$}

\def\hii{H\textsc{ii}}
\def\msol{M$_\odot$}
\def\hi{H\textsc{i}}
\def\2{$^{12}$CO}

\usepackage{graphicx}	
\usepackage{tikz}
\usepackage{amsmath}	
\usepackage{amssymb}	

\title[Linking stellar clusters]{Studying the interstellar medium to look for relics of triggered star formation among stellar clusters}

\author[S. Paron et al.]{
S. Paron,$^{1}$\thanks{E-mail: sparon@iafe.uba.ar }
A. Granada,$^{2}$
and M. B. Areal$^{1}$
\\
$^{1}$CONICET-Universidad de Buenos Aires. Instituto de Astronomía y Física del Espacio CC 67, Suc. 28, 1428 Buenos Aires, Argentina\\
$^{2}$ 
Centro Interdisciplinario de Telecomunicaciones, Electrónica, Computación y Ciencia Aplicada (CITECCA), Sede Andina, Universidad Nacional de
Río Negro, \\ Anasagasti 1463, San Carlos de Bariloche, R8400AHN Río Negro, Argentina\\
}

\date{Accepted XXX. Received YYY; in original form ZZZ}

\pubyear{2020}

\begin{document}
\label{firstpage}
\pagerange{\pageref{firstpage}--\pageref{lastpage}}
\maketitle

\begin{abstract}

Evidence of triggered star formation at large spatial scales involving stellar clusters is scarce.
We investigate a Galactic region ($l=130\fdg0, b=0\fdg35$) populated by several open stellar clusters that according to the last GAIA data release, are located at a distance of about 2.9 kpc. 
By analyzing the interstellar medium (ISM) at infrared, centimeter, and millimeter wavelengths towards this group of clusters we discovered a shell of material of about 2\d~in size at the same distance. We suggest that the shell, mainly observed at 12 $\mu$m and in the \hi~emission at 21 cm, was generated by the action of massive stars belonging to clusters Berkeley 7 and UBC 414, that lie at its center. 
Five clusters (MWSC0152, Czernik 6, Czernik 7, Berkeley 6, NGC 663, and NGC 654) lie at the border of this shell. 
From the comparison between the dynamical time of the discovered \hi~shell and the analysis of the ages of stellar populations in these clusters, 
we conclude that the expansion of the shell could have triggered in the past the formation of stars in some of them.
We point out that in order to find physical evidence supporting a genetic connection between stellar clusters, it is necessary not only to study the individual clusters and their stellar populations, but also to investigate their surrounding ISM at a large spatial scale.

\end{abstract}




\section{Introduction}

Triggered star formation has been extensively studied in the past decades and it was mainly proved at the interfaces between \hii~regions and molecular clouds. 
Populations of new stars are usually observed in molecular clouds that surround the photodissociation regions, showing that the expansion and the radiative action of the \hii~regions can trigger the formation of new stars (e.g. \citealt{figueira20,duro17,deha15,zav10,zav07,hoso05}). These processes can be studied towards relatively young regions, with at most a few 10$^{5}$ yr, and within spatial scales between 1 to 10 parsecs.

Triggered star formation due to cloud-cloud collisions has received special attention since some years ago, and evidences supporting such process have been found (e.g. \citealt{fukui21,dobbs20}, and references therein). The magnetic field amplified by the shock compression in the cloud-cloud collision can explain massive-cores and 
star formation in the collision interface \citep{inoue13,inoue18}. This process can take place at different spatial scales, 
from that of the typical \hii~bubbles to larger spatial scales ranging from a few tens to hundreds of parsecs, and the involved time scales in the formation of massive stars can be within 1 Myr after the shock \citep{inoue18,fukui18}.  

When the spatial and temporal scales become even larger, 
studying triggered star formation events becomes changelling. 
It is known that large star-forming complexes
produce stellar associations and star clusters. The pressure of young stars trigger gas accumulation on the periphery of cleared cavities in the interstellar medium (ISM), which often take the form of rings where new groups of stars form \citep{elme12}. At large spatial scales, this kind of rings are usually observed as atomic shells or supershells with sizes that go from tens to hundreds of parsecs, and even kiloparsecs \citep{mcclure02}. They can contribute to enhance the amount of molecular matter in the volumes affected by their expansion as theoretically proposed by for example \citet{inoue09} and \citet{heitsch08}, and new stars, or young clusters, can form from this molecular matter \citep{dawson11}. Over time, the stellar population of these clusters evolves, and the parental molecular clouds disperse \citep{fukui99}. Hence, looking for hierarchical triggered star formation at large spatial scales, and considering times scales ranging from some tens to hundreds of Myr, typical of intermediate-age open clusters, is challenging, and thus, it is necessary to make efforts in order to find causal connections among clusters and clusters complexes (e.g. \citealt{delafuente09}). This kind of studies are important because connecting such processes at large spatial and time scales allows us to better understand the dynamics and evolution of the Galaxy.

Evidence of triggered star formation among  evolved or intermediate-age stellar clusters is rare, indeed.
Besides studying the spatial location, distances, dynamics, and ages of such stellar clusters, we argue that it is mandatory to study the ISM around them with the aim to find pieces of physical evidence that suggest a present or an old connection that gives support, or at least allows us to infer, the presence of relics of triggered star-forming processes. In the following sections, based on the study of the ISM, cluster parameters and their stellar populations, we present observational evidence supporting a genetic relation within a group of Galactic open clusters.

\section{Presentation of the case}

\begin{table}
\caption{Parameters of the stellar clusters lying in the analyzed region.}
\centering
\tabcolsep=0.12cm
\begin{tabular}{lccccccc}
\hline\hline
Cluster      &   RA     &   Dec.   & \multicolumn{3}{c}{Distance [kpc]}   &  \multicolumn{2}{c}{log(t) [yr]}   \\
             &          &          &PGA$^{1}$ & CG$^{2}$ & K$^{3}$       &  CG & K        \\
\hline      
  Berkeley~7   & 28.546  & 62.366  & 2.9 & 2.9   & 2.6   & 7.53   & 7.25  \\
  UBC~414      & 28.513  & 61.992  & 2.9 & 2.9   & --    & 7.47   & --   \\ 
  IC~166       & 28.094  & 61.857  & 5.1 & 5.3   & 4.8   & 9.12   & 9.00  \\
  COIN-Gaia32 & 28.194  & 63.066  & 1.3 & 1.2   & --    &  8.56  & --    \\ 
  Berkeley~5   & 26.928  & 62.935  &  -  & 7.1   & 6.0   & 9.06   & 8.83  \\
  Czernik~5    & 28.927  & 61.355  & 3.9 & 4.1   & 2.4   & 8.13   & 8.61  \\ 
  Czernik~6    & 30.543  & 62.838  & 2.9 & 2.5   & 1.3   & 7.45   & 7.98  \\ 
  COIN-Gaia33 & 30.276  & 61.475  & 6.0 & --    &  --   & --     & --   \\
  NGC~654      & 26.008  & 61.883  & 3.1 & 3.1   & 1.8   & 6.99   & 7.25  \\ 
  SAI~16       & 31.366  & 62.265  &  -  & 4.9   & 4.7   & 9.00   & 9.15  \\ 
  Berkeley~6   & 27.803  & 61.061  & 3.1 & 3.0   & 2.5   & 8.35   & 8.60  \\
  COIN-Gaia34 & 31.231  & 61.776  & 1.0 & 1.0   & --    & 8.34   &  --   \\
  NGC~663      & 26.586  & 61.212  & 2.9 & 3.0   & 2.1   & 7.47   & 7.50  \\
  Czernik~7    & 30.754  & 62.253  & 3.6 & --    & 3.0   &  --    & 9.12  \\
  MWSC0152     & 29.332  & 63.490  & 3.1 & --    & 2.4   &  --    & 6.60  \\
\hline
\multicolumn{3}{l}{$^{1}$ PGA: This work }\\  
\multicolumn{3}{l}{$^{2}$ CG: \citet{can18,can20} } \\
\multicolumn{3}{l}{$^{3}$ K13: \citet{khar13} } \\
\end{tabular}
\label{tabpresent}
\end{table}

We focus on a Galactic region centered at $l=130\fdg0, b=0\fdg35$ with a radius of 1\fdg5, that according to 
\citet[e.g.][]{khar13} or  \citet{can18,can20} is populated by several
open stellar clusters. Table\,\ref{tabpresent} presents the coordinates, distances, and ages corresponding to these clusters. Following the mentioned authors, we find that most of them are located at a distance lying in the range of about 2.4--3.1 kpc (see columns labelled as CG and K), suggesting that they could be located close enough to be physically related. 

Given that the GAIA Early Data Release 3 (GAIA EDR3, \citealt{gaia3}) has been recently made available, for each of the clusters in Table\,\ref{tabpresent}, we consider the cluster members given by \citet{can20}, when available, or by \citet{khar13}, and calculated the mean value of GAIA EDR3 parallaxes, considering those stars with G brighter than 15mag, which ensures small parallax errors \citep{gaia3}. Using these values we derive a new mean distance to the clusters, tabulated in column labelled as PGA in Table\,\ref{tabpresent}. For two clusters, Berkeley~5 and SAI~16, the quality of the parallaxes does not allow to provide a reliable distance, but their previous determinations in the literature (see columns CG and K in Table\,\ref{tabpresent}) are certainly larger than 4.5 kpc. We can see that while ICC 166, Czernik~5, and COIN-Gaia~33 have distances larger than 3.5 kpc, COIN-Gaia~32 and COIN-Gaia~34 are located closer than 1.5 kpc. The remaining seven clusters, Berkeley 7, UBC~414, NGC~654, Berkeley~6, NGC~663, Czernik~6 and  MWSC152, have a distance between 2.9 kpc and 3.1 kpc, and Czernik~7 could be considered to be close to 3 kpc.

Thus, we decide to 
investigate the ISM at a large spatial scale towards this complex in order to search for 
some evidence pointing to a genetically connection among the stars belonging to these clusters.

\section{Data}

To study the ISM towards this region populated of stellar clusters we analyzed data from three large-scale databases. We used IR data 
extracted from the Wide-field Infrared Survey Explorer (WISE; \citealt{wright10}). To study the distribution of cold
gas towards the region we used \hi~data retrieved from the Canadian Galactic Plane Survey (CGPS; \citealt{taylor03}), which have an angular and spectral resolutions of 1\m, and 1.3 \ks, respectively, and \2 J=1--0 emission obtained
from Five College Radio Astronomical Observatory (FCRAO) CO Survey of the Outer
Galaxy \citep{heyer98} with 45\s~and 0.8 \ks~in angular and spectral resolution.

\section{Results}

In the following sections we present the results obtained from each dataset, with the aim of investigating
a physical connection among some of the mentioned stellar clusters.

\subsection{IR emission}

\begin{figure}
    \centering
	\includegraphics[width=8.4cm]{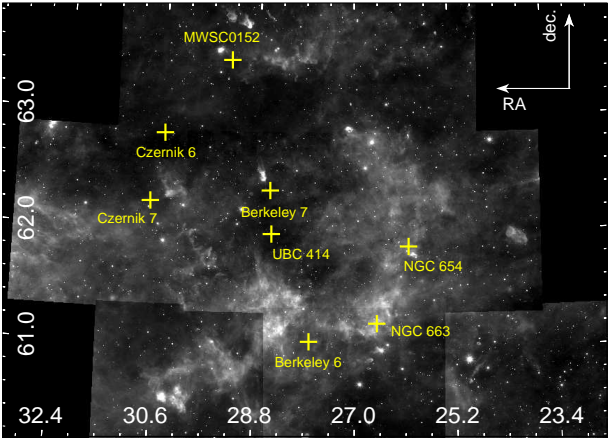}
    \caption{Mosaic image displaying the 12 $\mu$m emission obtained from WISE. The positions of the clusters are indicated with crosses.}
    \label{figir12}
\end{figure}

Figure\,\ref{figir12} shows the 12 $\mu$m emission towards the analyzed region conformed by eight mosaics obtained from the WISE database. Stellar clusters in the distance range of about 2.9 to 3.1 kpc are included. A shell-like structure with an elliptical morphology, centered close to the clusters Berkeley~7 and UBC 414 can be observed. The clusters NGC~654, NGC~633, and Berkeley~6 are located
over the bulk of IR emission that constitutes the south-southwest portion of the shell, while clusters Czernik~7, Czernik~6, and MWS0152 lie towards the north-northeast border which is more diffuse. 
This shell has many interesting features (some of them are discussed below), and its emission at 12 $\mu$m should be due to the presence of polycyclic aromatic hydrocarbons (PAHs) and dust grains (e.g. \citealt{reach17}), that absorbs UV photons and are heated by the radiation of massive stars. 
This scenario is compatible with the fact that Berkeley~7 and/or UBC~414
has/have emptied the ISM at its surroundings through the radiation and winds of its stellar members, and likely generated the observed IR shell. The expansion of this shell could have affected some of the clusters located at its border. This issue is studied in the following sections.

\begin{figure}
    \centering
	\includegraphics[width=8.4cm]{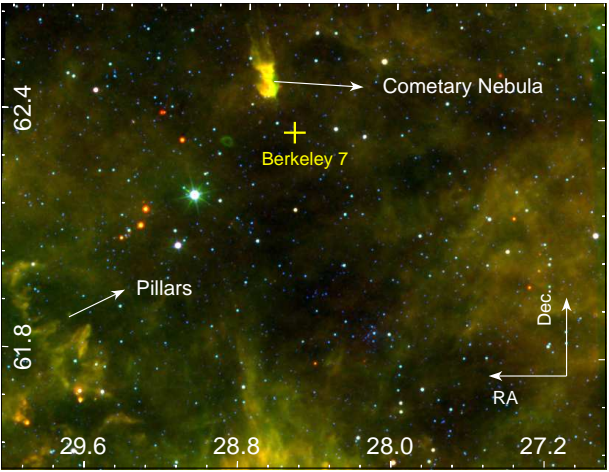}
    \caption{Three-colour image displaying the 22, 12, and 4.6 $\mu$m emission obtained from WISE in red, green, and blue, respectively. Two interesting IR features are remarked.}
    \label{figirCol}
\end{figure}

Figure\,\ref{figirCol} presents a WISE three-colour image remarking two conspicuous structures as seen at IR: a complex of pillar-like features towards the southeast of Berkeley~7, and a cometary nebula close to the same cluster. It is well known that this kind of structures are generated by the action of massive stars (e.g. \citealt{mook19,dju17,gahm06}), which through their ionizing and photodissociation shock fronts compress and shape the material that surrounds them. Dense and dusty molecular clumps are usually shaped in pillars of gas pointing towards the ionizing sources, and in globules detached from the parental molecular cloud. Cometary globules are small and dense clouds consisting of a dense head, surrounded by a bright rim, and prolonged by a diffuse tail. Thus, the features remarked in Fig.\,\ref{figirCol} are evidence supporting the presence of massive stars in the region. 

\subsection{Molecular gas}

\begin{figure}
    \centering
	\includegraphics[width=8.3cm]{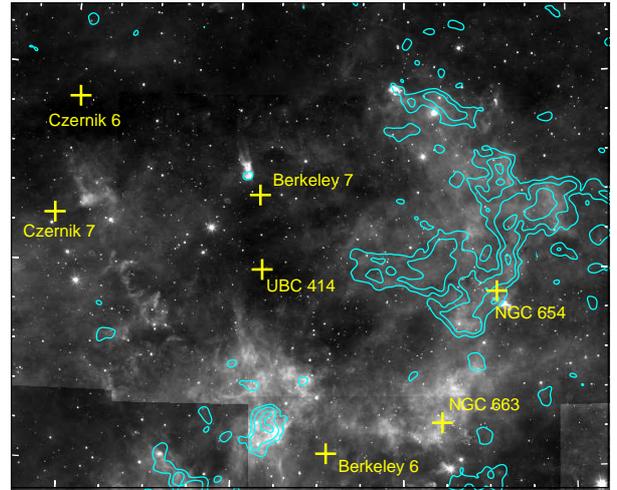}
    \caption{IR emission at 12 $\mu$m with contours of the \2 J=1--0 integrated between $-27$ and $-42$ \ks. The contours levels are 2, 6, and 15 K \ks. The positions of the clusters are indicated with crosses.}
    \label{figCO}
\end{figure}

\begin{figure}
    \centering
	\includegraphics[width=8.4cm]{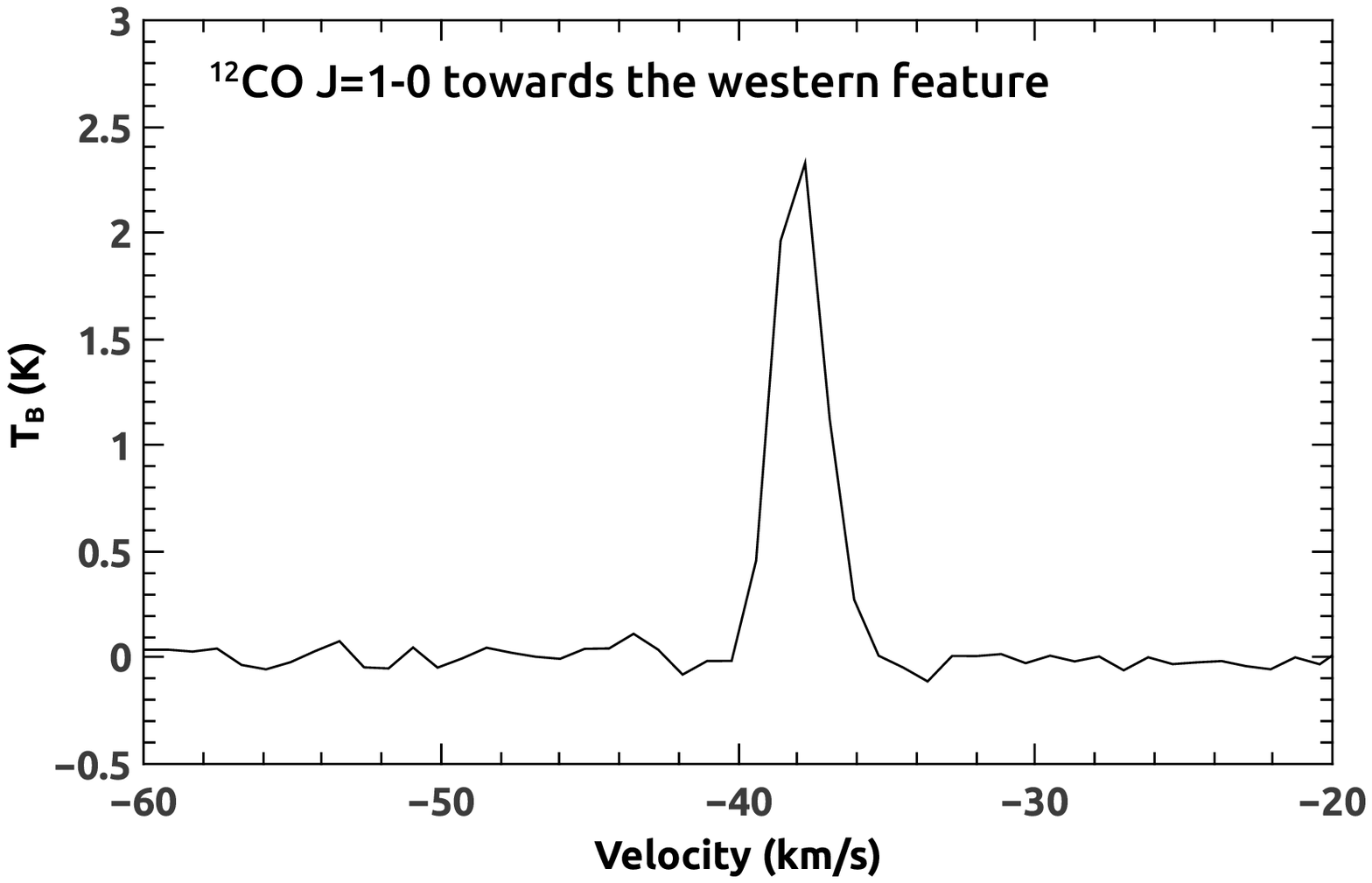}
    \caption{\2 spectrum obtained towards the western molecular feature at RA 27\fdg356, dec. 62\fdg002.}
    \label{spect}
\end{figure}

Regarding the molecular gas distribution, we inspected the CO data cube along the whole velocity
range and we found some CO emission likely related to the shell observed at IR in the velocity range
that goes from $-27$ to $-42$ \ks~(see Fig.\ref{figCO}). The CO spectra corresponding to the western molecular feature peak at about $-37$ \ks~(see Fig.\ref{spect}). Using the rotation curve of \citet{reid14}, this velocity yields a distance of about 2.8 kpc for the observed molecular cloud in perfect agreement with the distance range in which the stellar clusters lie.

\begin{figure}
    \centering
	\includegraphics[width=8.4cm]{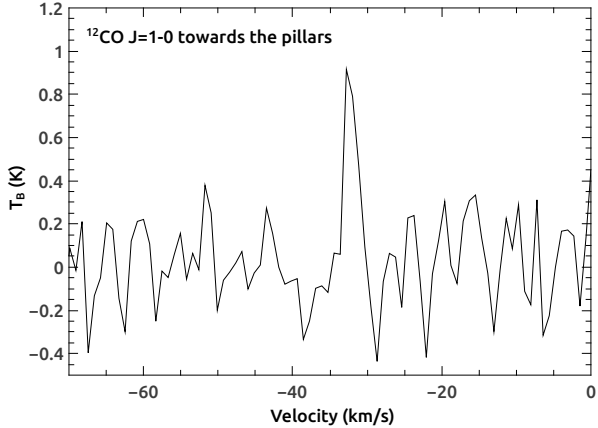}
    \caption{\2 spectrum obtained towards the most extended pillar-like feature, at RA 29\fdg644, dec. 61\fdg835. The marginal signal at $-33$ \ks, is almost at 3$\sigma$ the rms noise.}
    \label{COpillar}
\end{figure}

Figure\,\ref{figCO} shows that the observed molecular clouds have a good
morphological correspondence to the IR emission towards the west of the shell. 
In addition, a molecular fragment is observed eastwards Berkeley~6, and a small molecular clump appears exactly at the head of the cometary nebula described above. This integrated \2 map shows that the pillar-like features lack of molecular gas. However, by inspecting the spectra corresponding to this region we find a weak emission 
at $-33$ \ks~(see Fig.\,\ref{COpillar}) related to the pillars. While this marginal signal suggests that the molecular gas was highly photodissociated in this region, its systemic velocity and the spatial association with the pillars show that they are indeed located at the same distance of the analyzed complex, and thus they should be sculpted by the action of the stars that have generated
the shell observed at the IR wavelengths.

The rest of the IR shell lacks of molecular gas. This is consistent with a molecular shell 
fragmented and partially destroyed by the radiation and winds from massive stars, and probably also by supernova explosions \citep{zub14}. The molecular feature lying towards the west may be result of the densest region of a molecular shell that was not completely destroyed yet. Moreover, the clumpiness in the observed molecular structures supports the scenario in which the fragmentation of a molecular cloud with a shell morphology leads the formation of stars and possibly stellar clusters  \citep{klessen01}. 

To characterize the most conspicuous molecular features we calculate the H$_2$ column density using the relation: N(H$_{2}) = X \int {\rm T(CO)} dv$ cm$^{-2}$, where $X = 1.9 \times 10^{20}$ (K \ks)$^{-1}$  \citep{digel95}, and the mass from: ${\rm M} = 2 m_{H} d^2 \Omega \sum_{i} {\rm N_i(H_2)}$, where $m_H$ is the H atom mass, $d = 2.8$ kpc, and $\Omega$ the beam area of the CO data. The summation was performed over the area delimited by the 2 K \ks~contour. Hence, the mass values obtained for the western molecular large feature and the small cloud observed between the clusters Berkeley~6 and Czernik~5 are $9.3 \times 10^{4}$ and  $ 1.0 \times 10^{4}$ \msol, respectively.  
Additionally, the mass of the small clump (1\m~in radius) related to the head of the cometary nebula was calculated in 170 \msol.

\subsection{Neutral atomic gas} 
\label{cocloud}

\begin{figure}
    \centering
	\includegraphics[width=8.3cm]{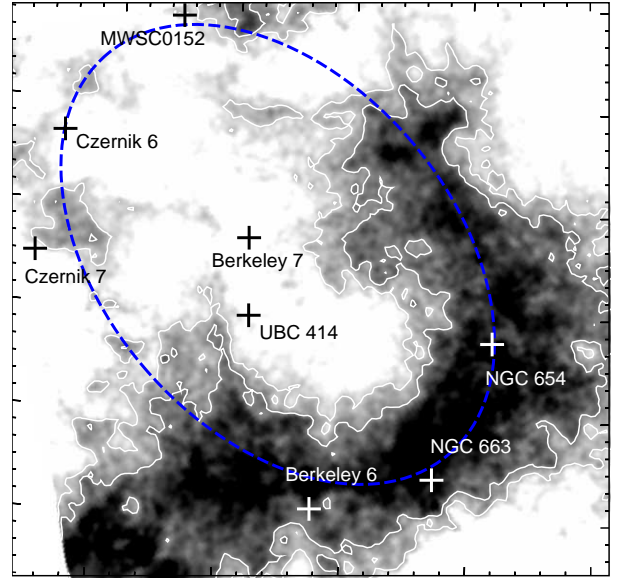}
    \caption{\hi~emission integrated between $-28$ and $-36$ \ks. The contour levels are 320 and 380 K \ks. The positions of the clusters are indicated. The blue dashed ellipse remarks the \hi~shell. }
    \label{figHI}
\end{figure}

After a careful inspection of the \hi~data cube we found an interesting structure extending from $-28$ to $-36$ \ks~(see Fig.\,\ref{figHI}).  
This shell-like \hi~feature with an elliptical morphology (semi-axes of 0\fdg8 and 1\fdg2) appears centered close
to the position of Berkeley~7 and has a perfect morphological correspondence with the structure observed at the IR emission, mainly towards the southwest. Indeed, this kind of \hi~features is a typical galactic structure classified as \hi~shell (e.g. \citealt{mcclure02,rel07}), which are massive objects, usually detected as
voids in the neutral hydrogen emission as is the case. It is known that they are formed through the combined
effects of stellar winds and supernovae, ionizing the neutral
medium and sweeping up a massive expanding shell. Thus, in this case, it is very likely that the \hi~shell was generated by the action of stars located at its interior, and probably, some, or most of them, belonging to Berkeley~7 and/or UBC~414. Moreover, the velocity range in which the \hi~shell extends is in agreement with that of the molecular features, showing that the \hi~shell is located at the same distance as the molecular clouds, in coincidence with the distance of the stellar clusters included in the figures.

\begin{figure*}
    \centering
	\includegraphics[width=4.3cm]{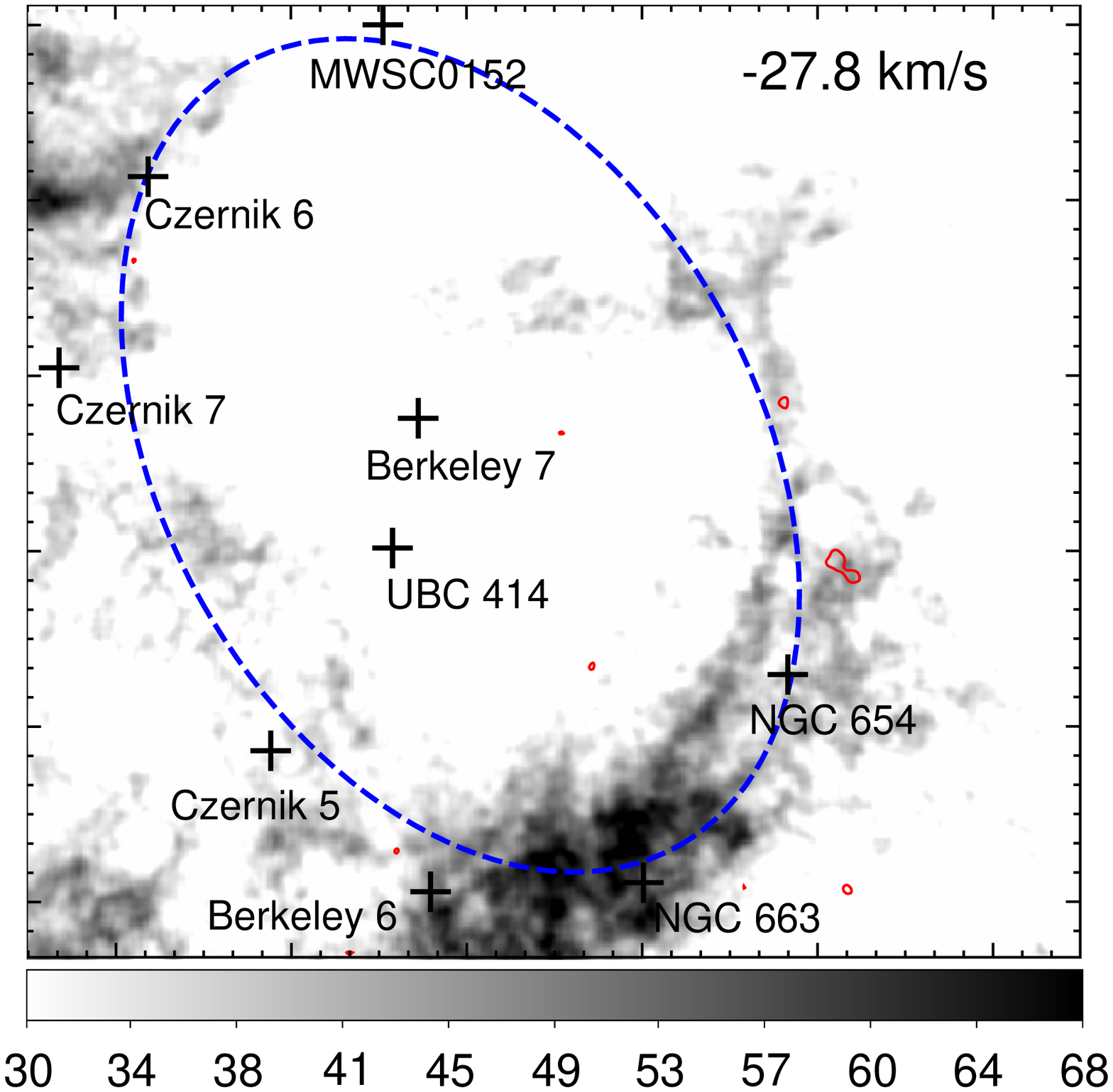}
	\includegraphics[width=4.3cm,trim=0 -2cm 0 0]{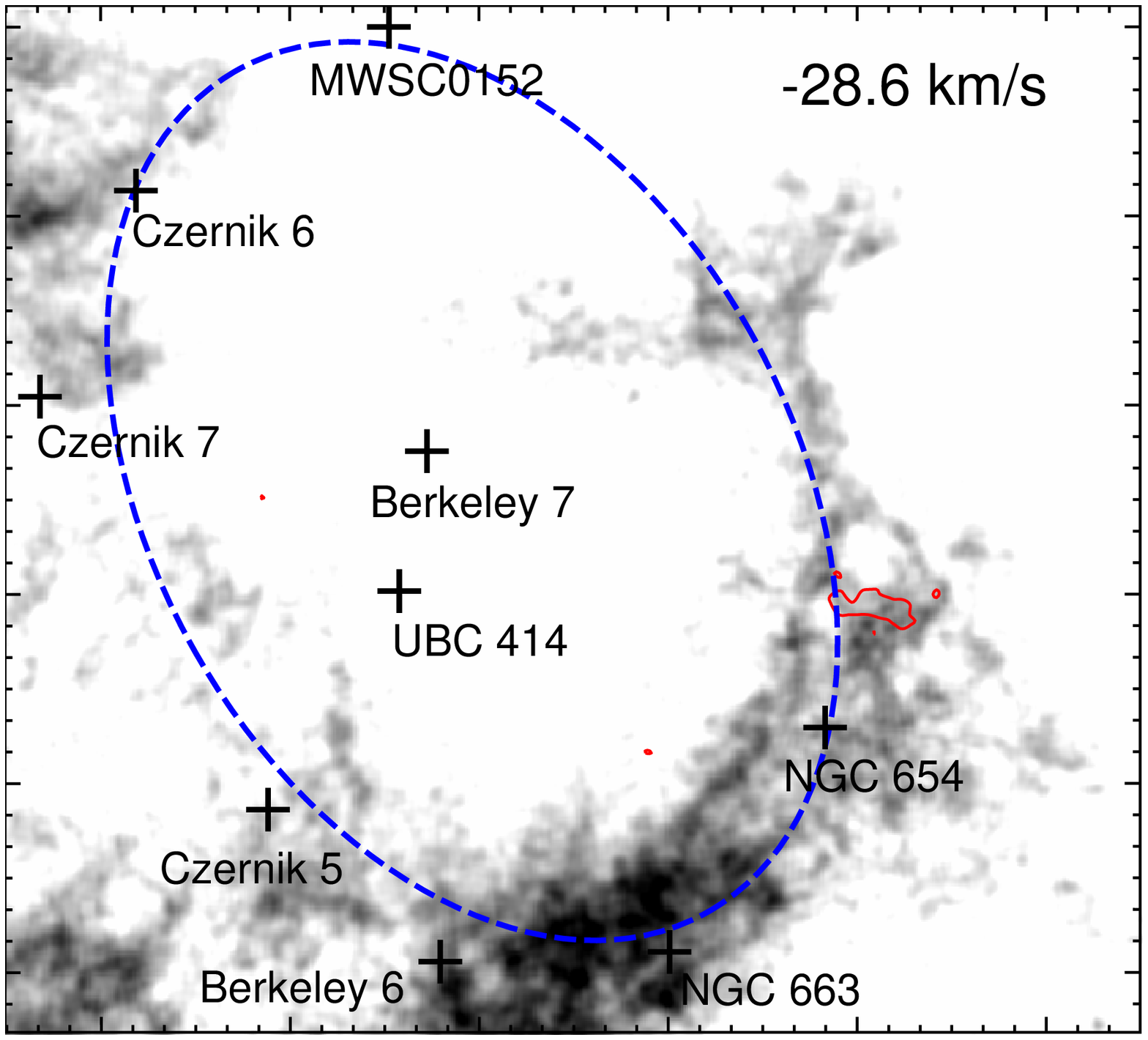}
    \includegraphics[width=4.3cm,trim=0 -2cm 0 0]{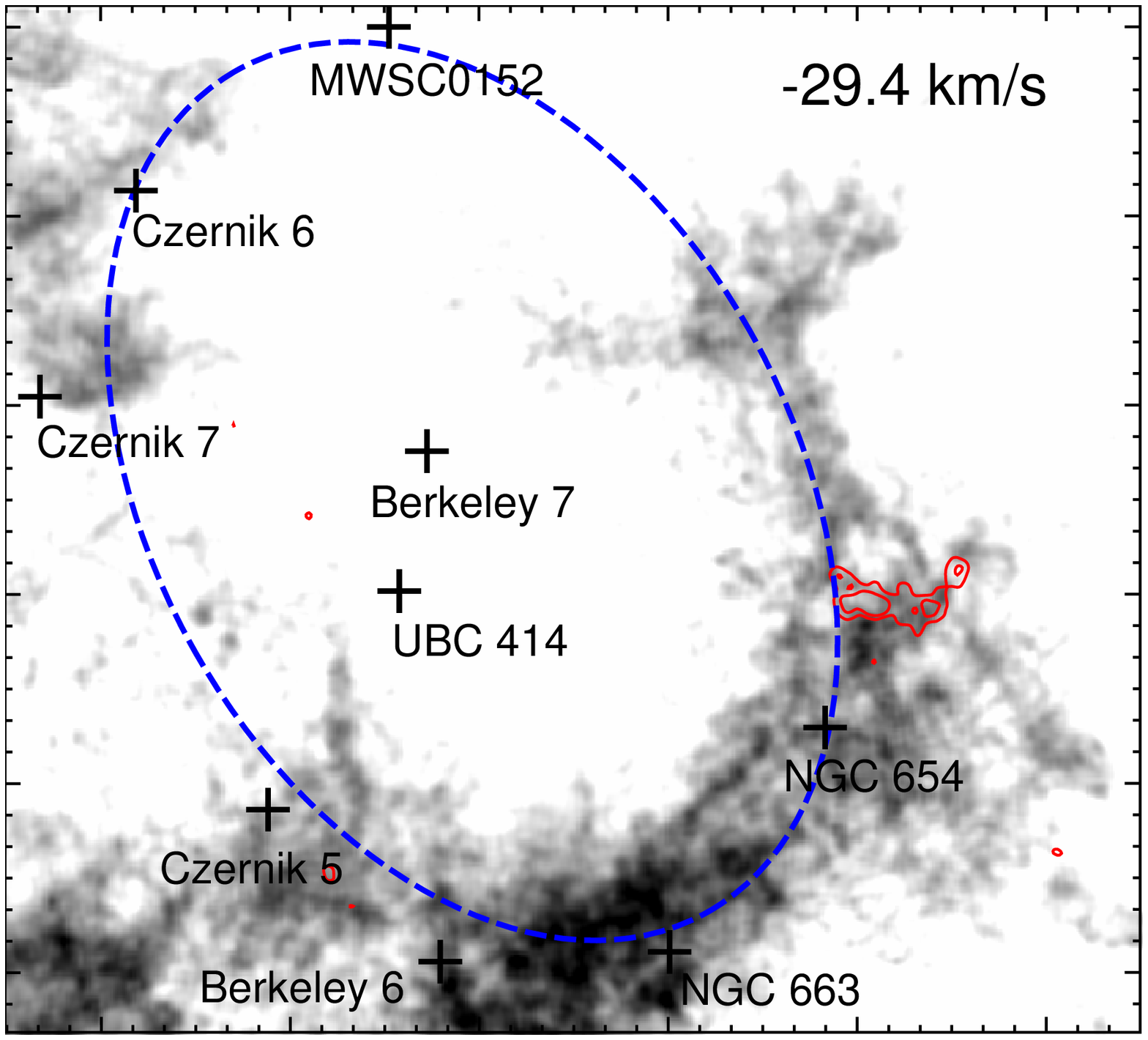}
	\includegraphics[width=4.3cm,trim=0 -2cm 0 0]{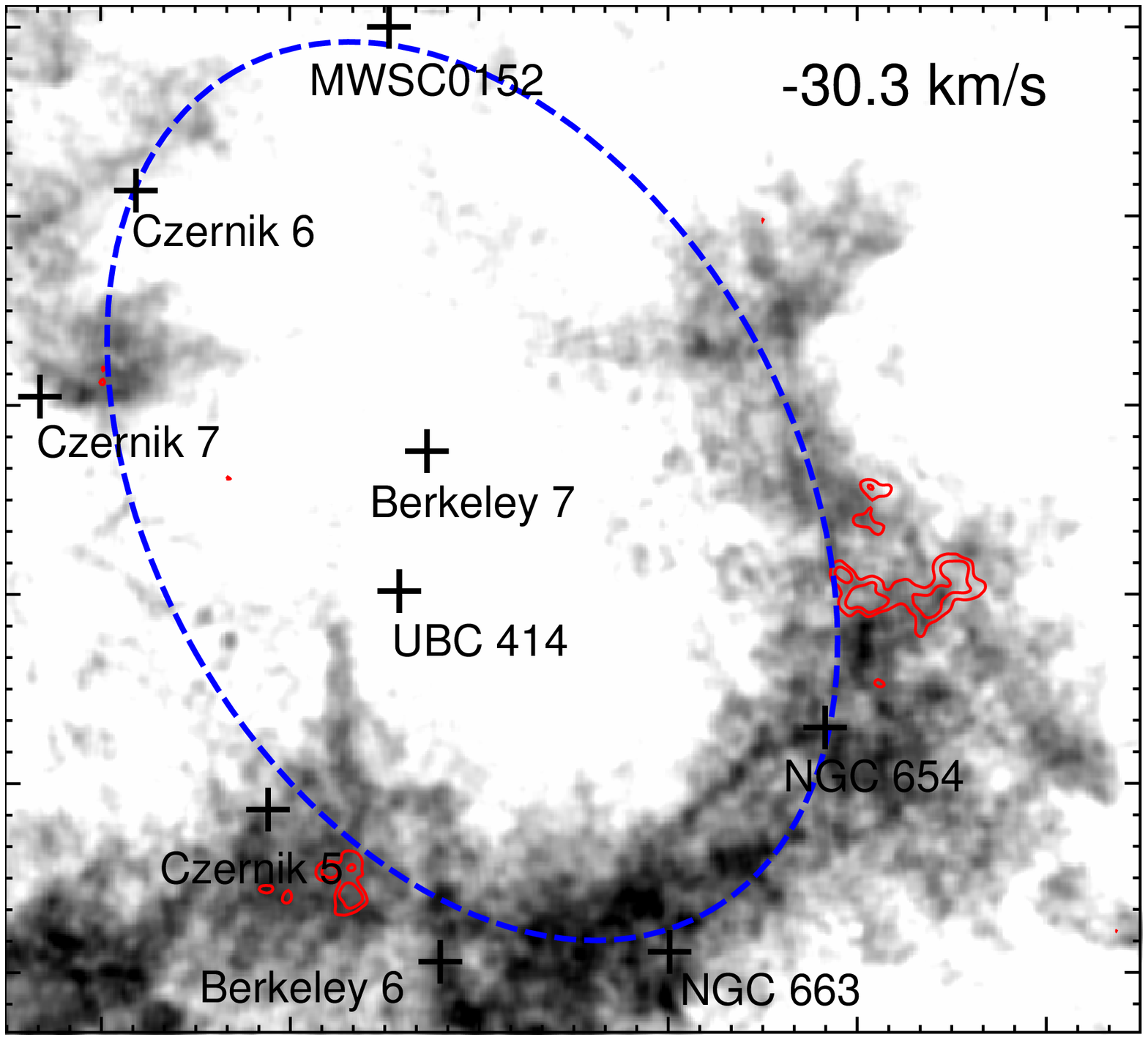}
	\includegraphics[width=4.3cm]{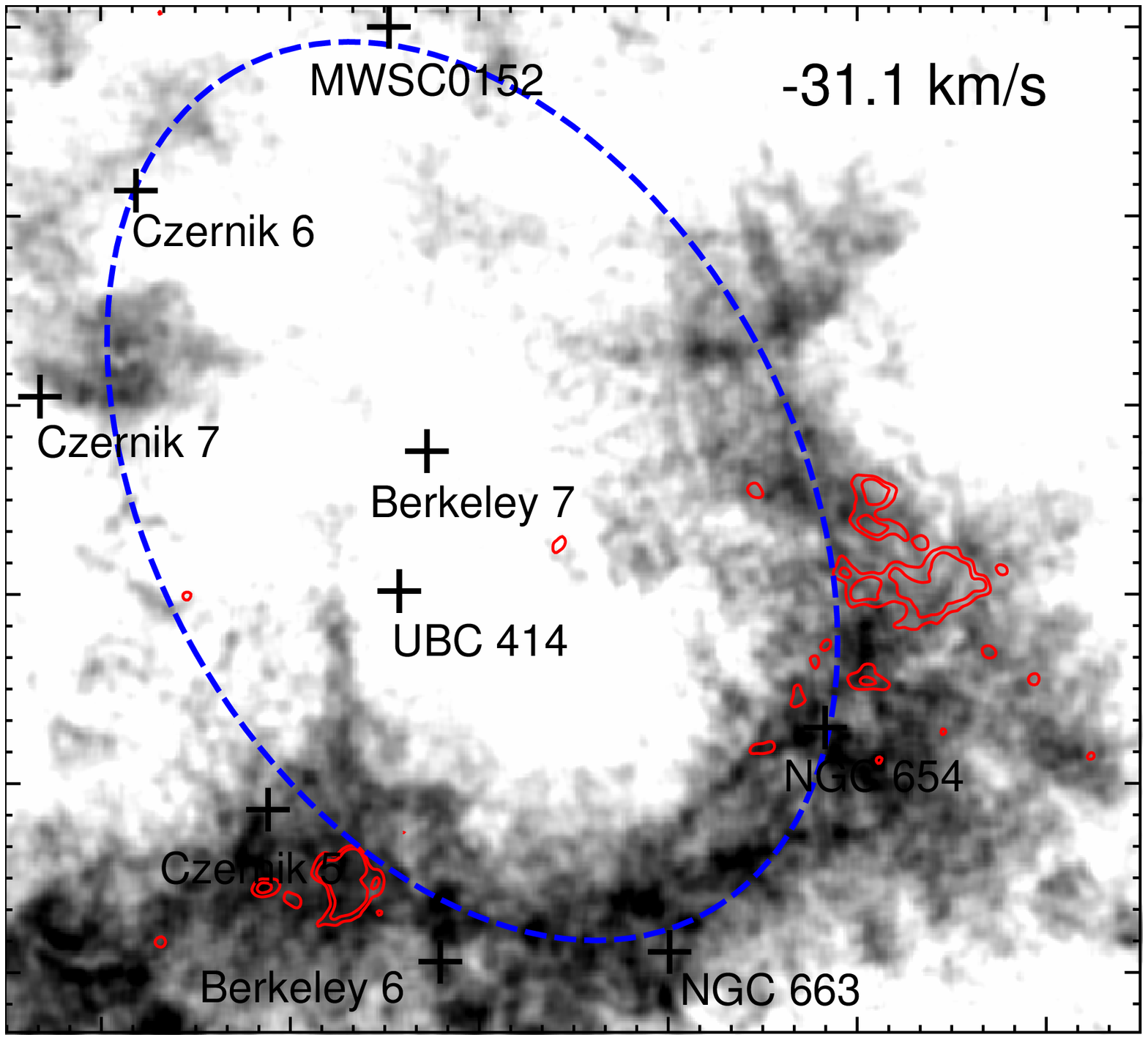}
	\includegraphics[width=4.3cm]{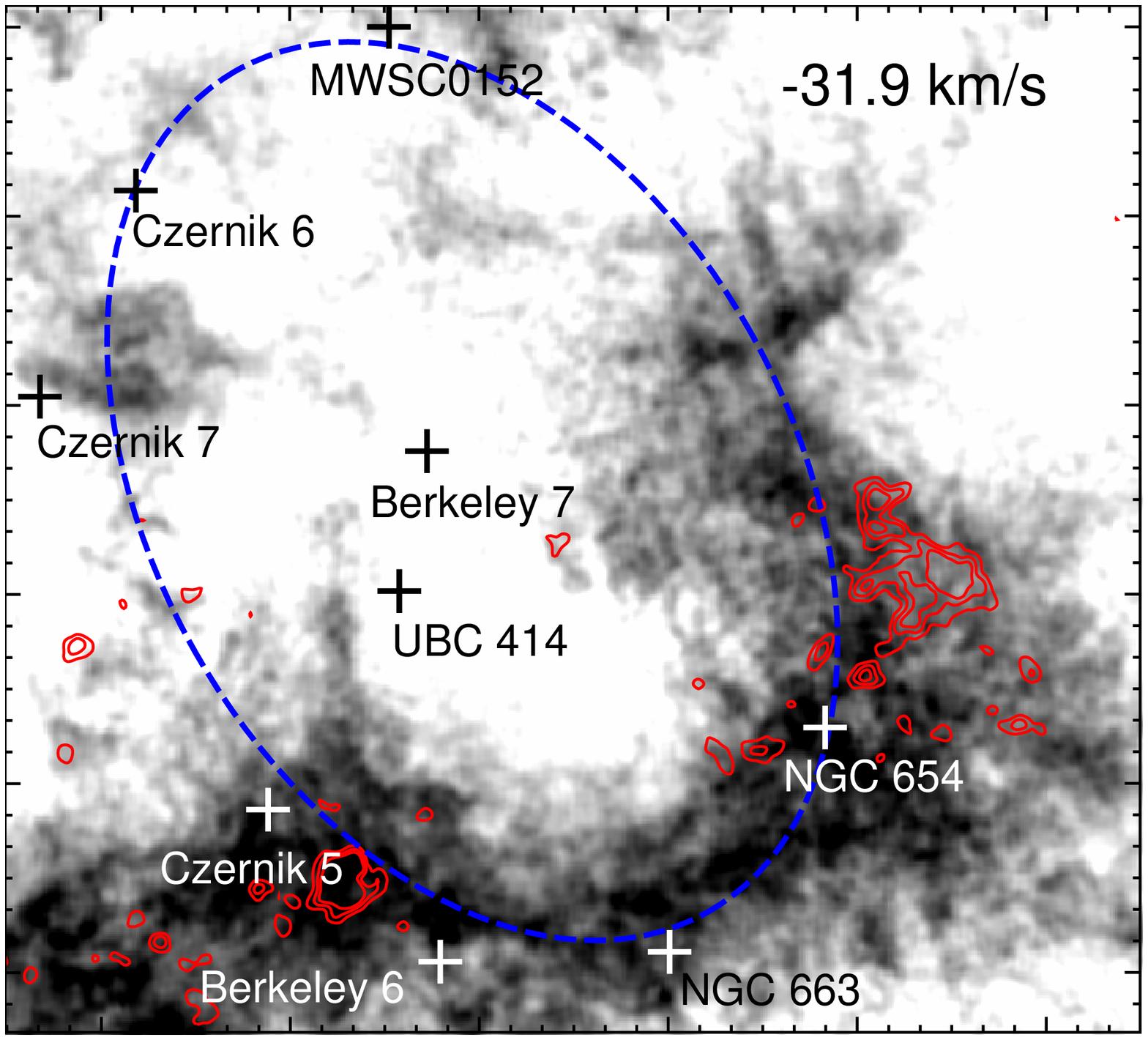}
	\includegraphics[width=4.3cm]{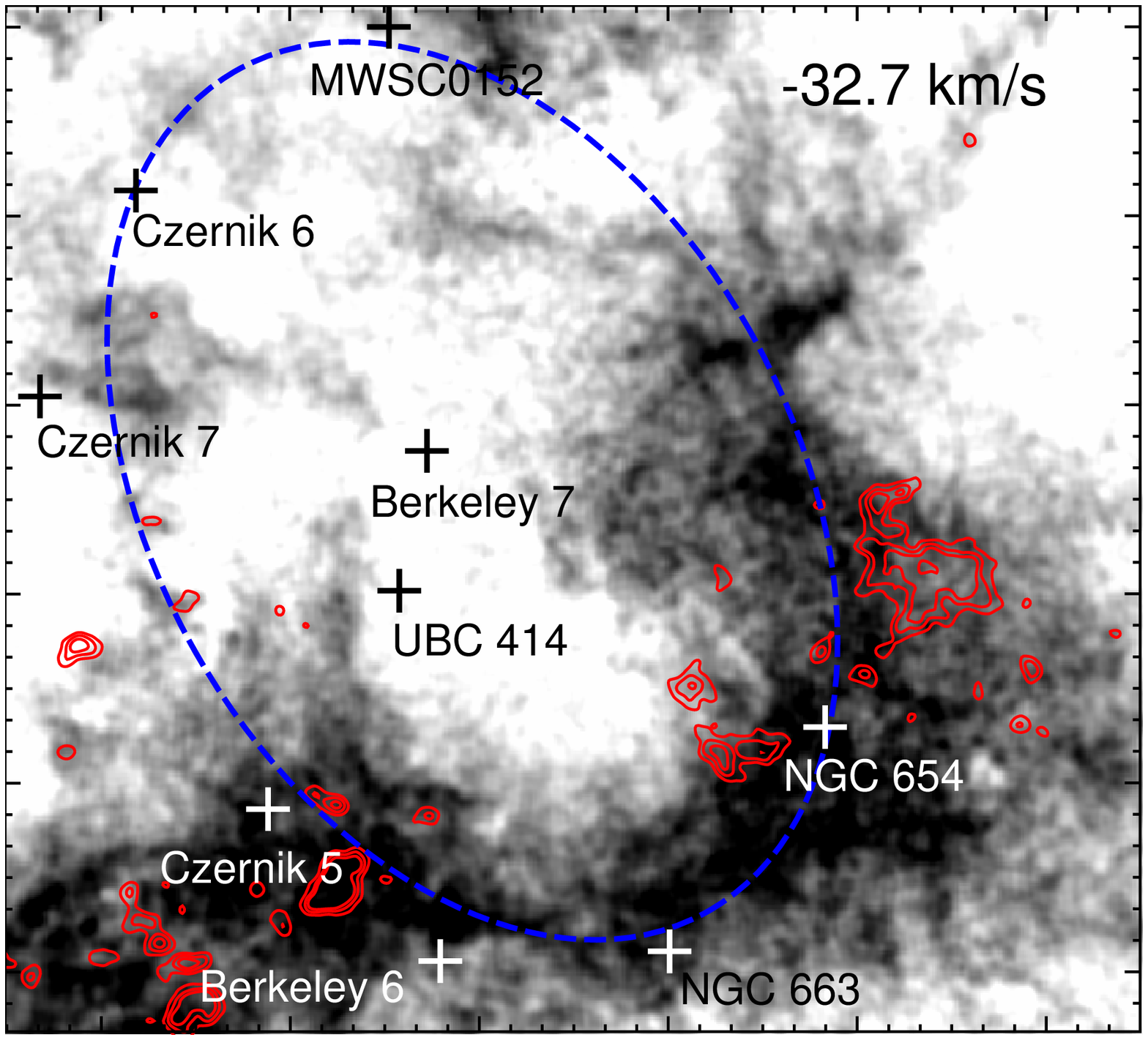}
	\includegraphics[width=4.3cm]{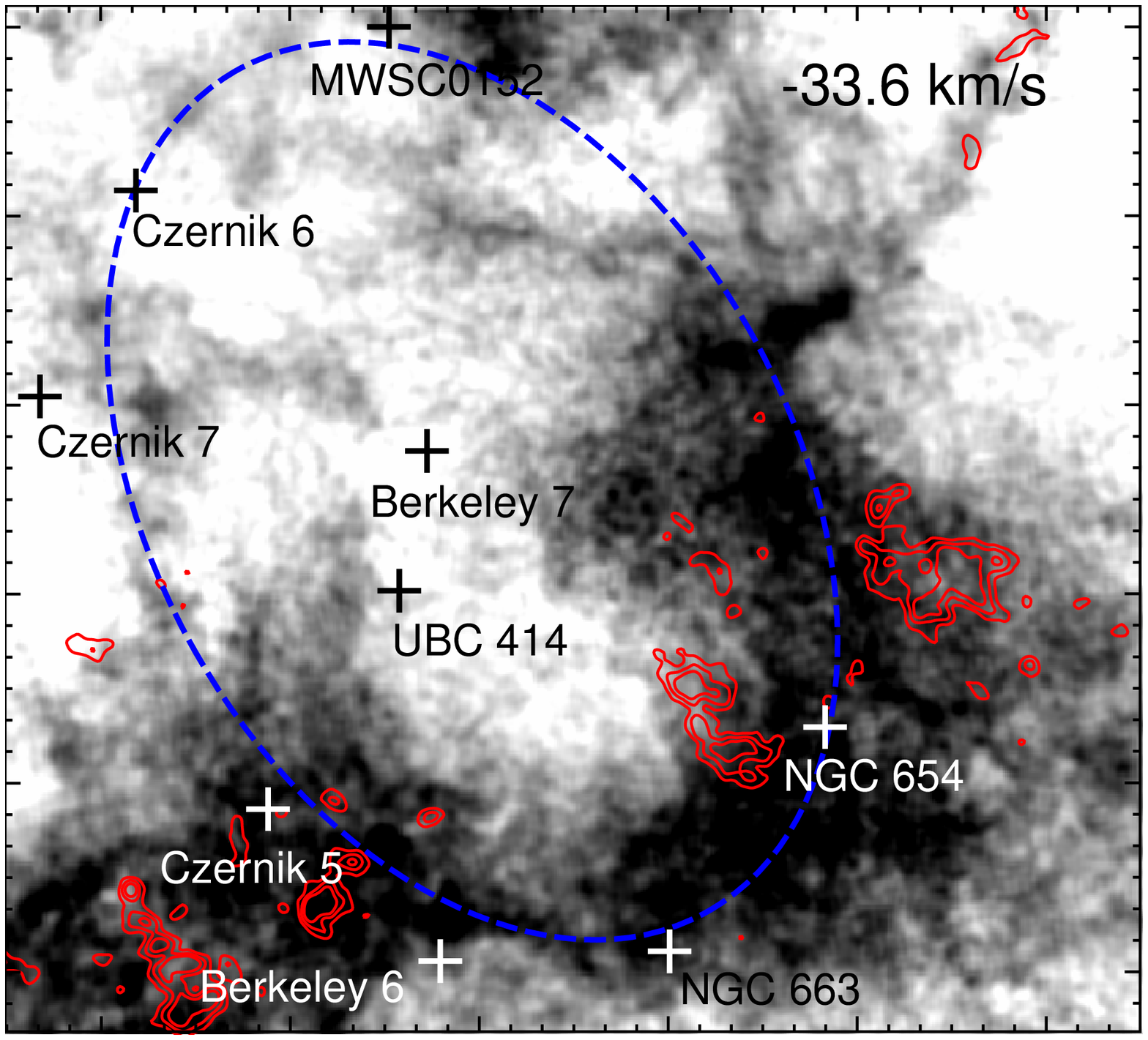}	
    \caption{Channel maps of the \hi~emission (grays) and the \2 J=1--0 emission (red contours) from $-27.8$ to $-33.6$ \ks~each $\sim0.8$ \ks. The colourbar is presented at the first panel and is in K, and the \2 contours levels are: 0.65, 1.50, 2.50, and 4.0 K. The positions of the clusters are indicated and the blue dashed ellipse remarks the \hi~shell. }
    \label{chann1}
\end{figure*}

\begin{figure*}
    \centering
	\includegraphics[width=4.3cm]{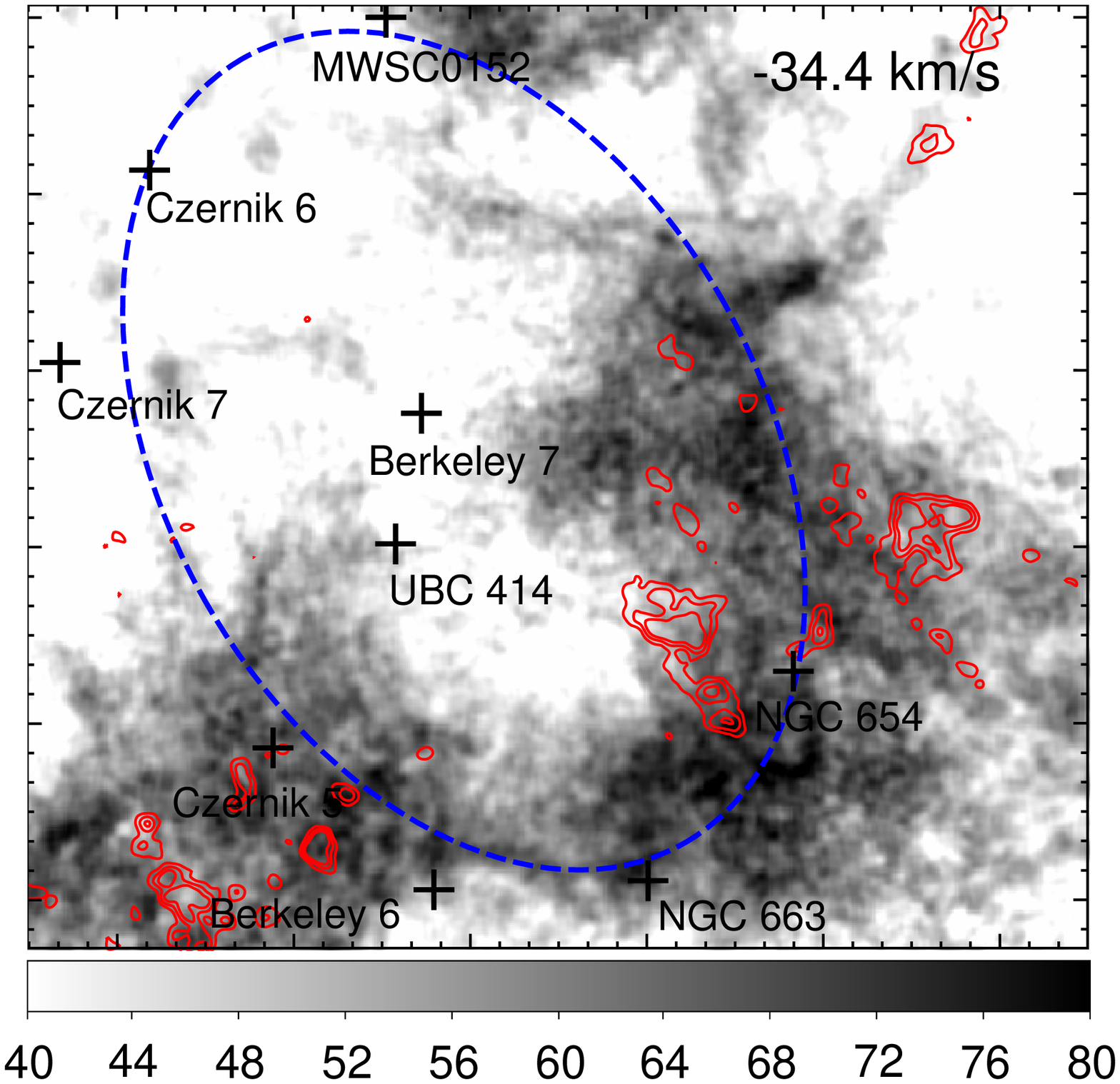}
	\includegraphics[width=4.3cm,trim=0 -2.3cm 0 0]{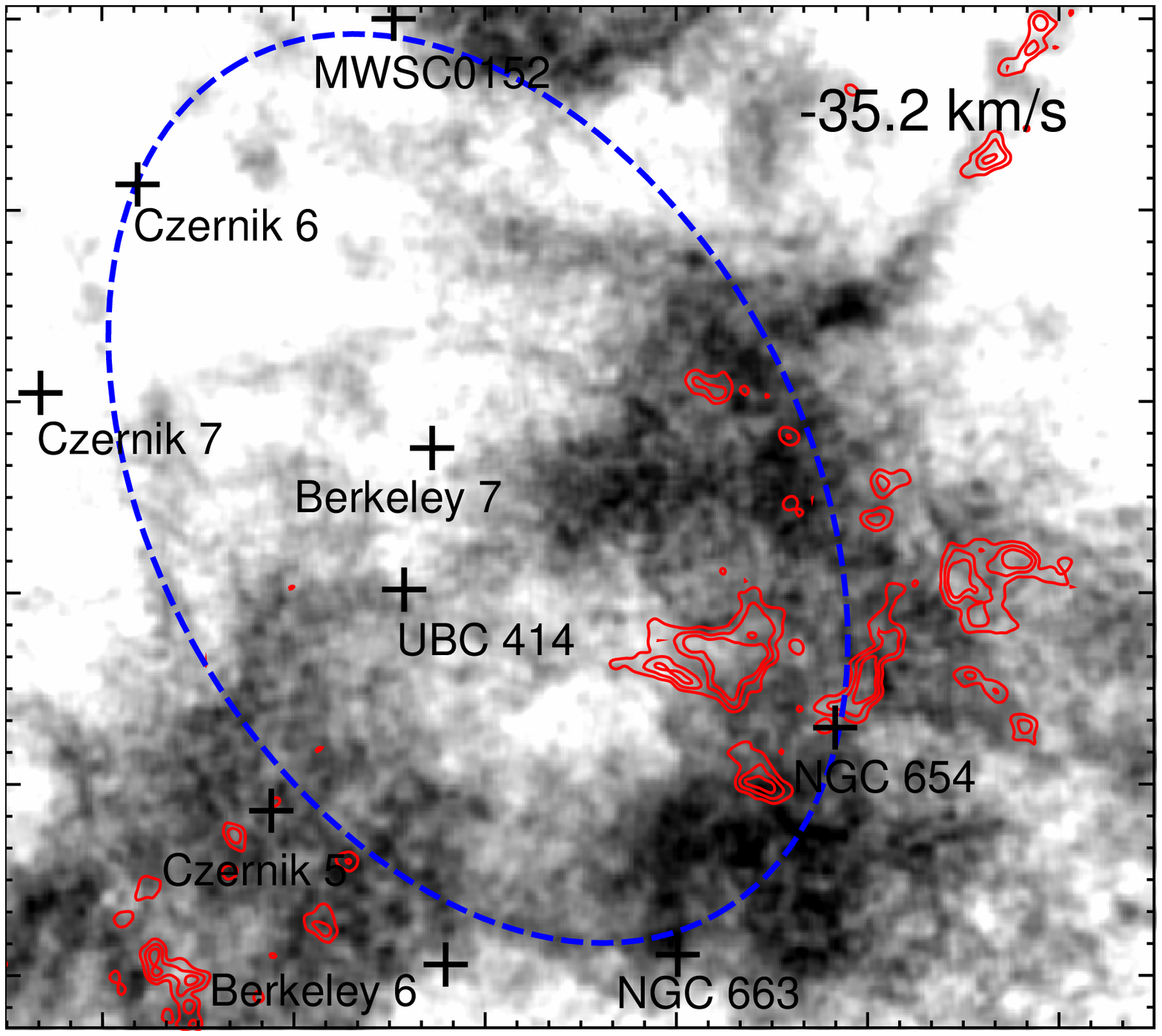}
    \includegraphics[width=4.3cm,trim=0 -2.3cm 0 0]{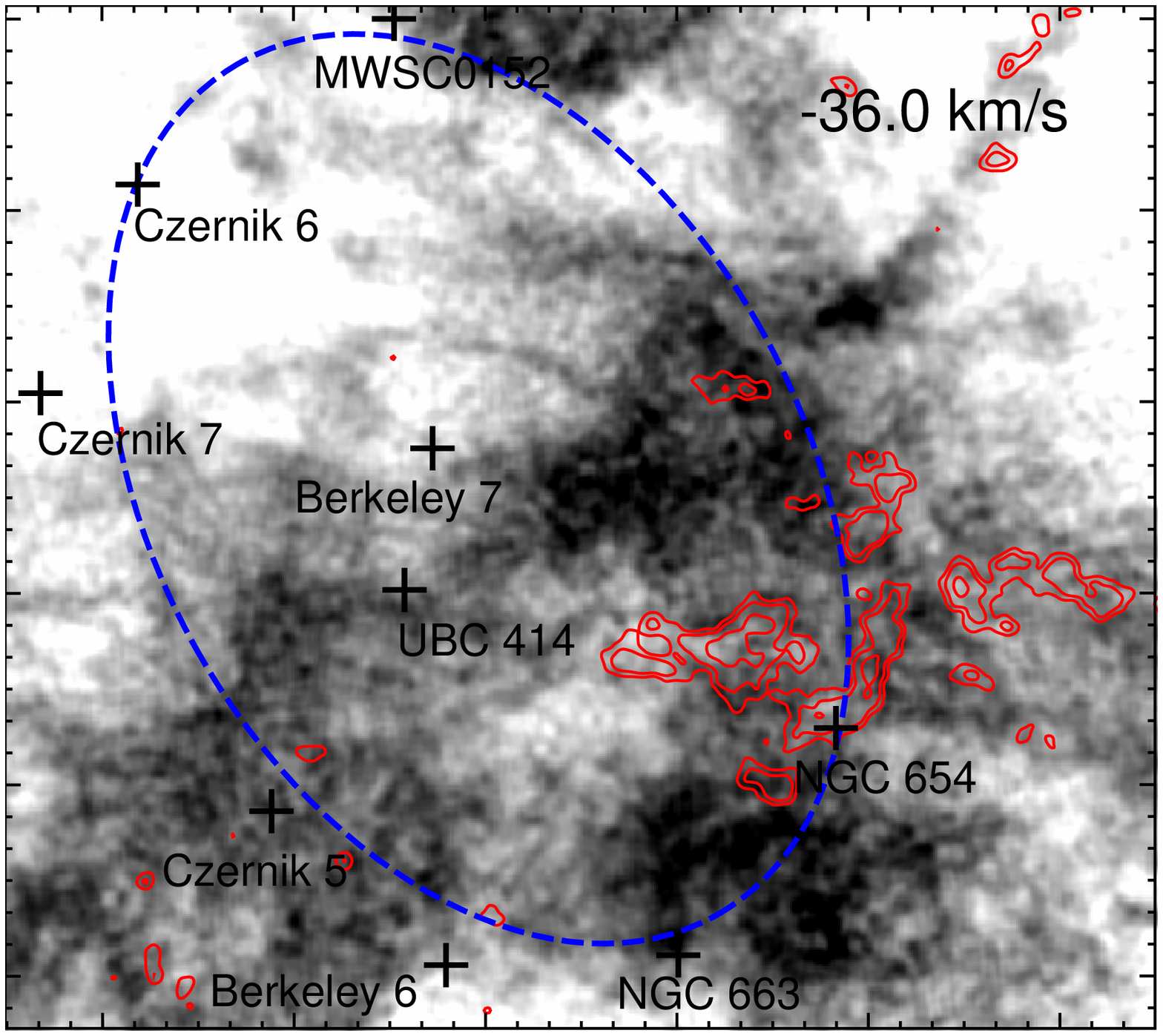}
	\includegraphics[width=4.3cm,trim=0 -2.3cm 0 0]{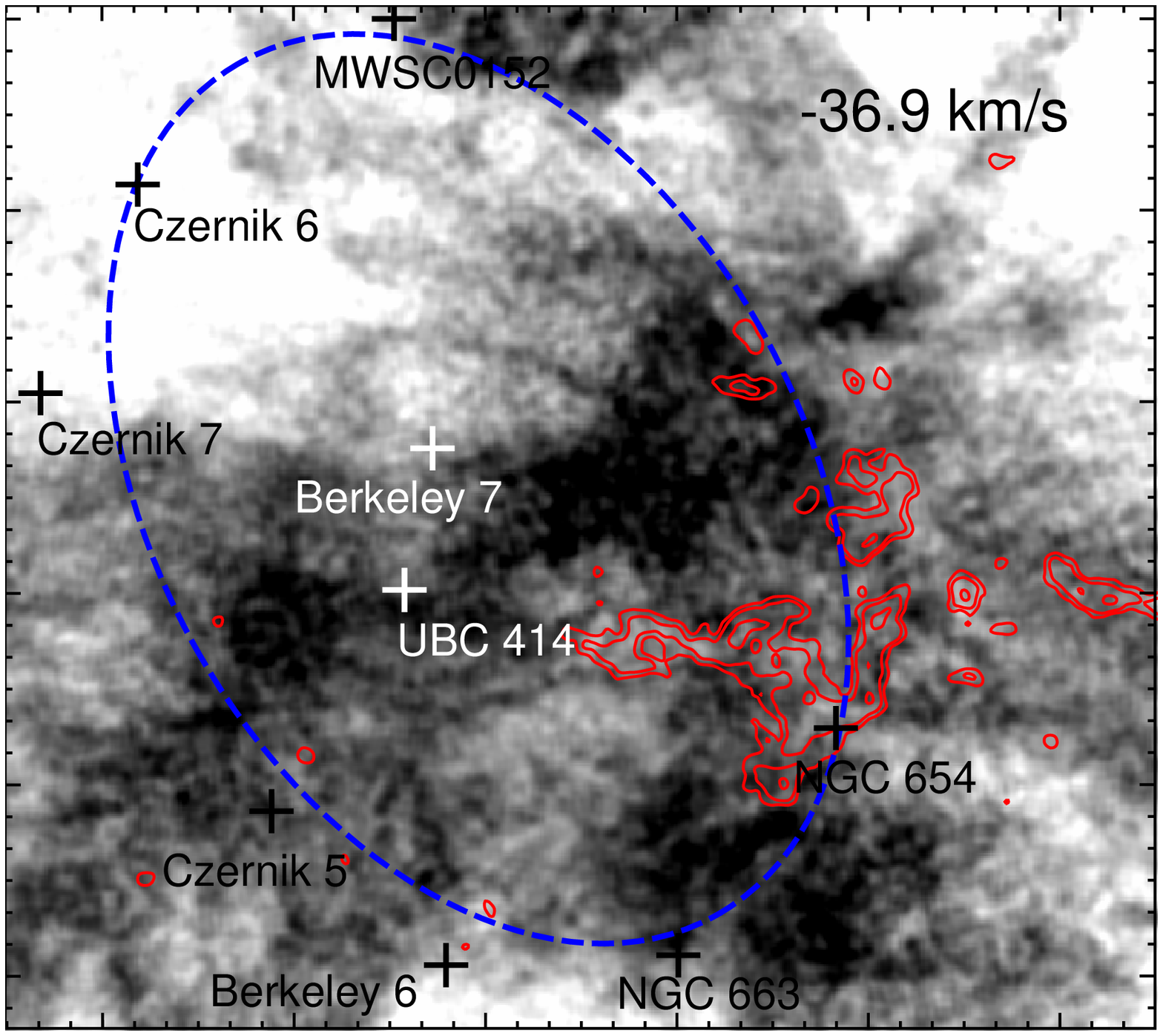}
	\includegraphics[width=4.3cm]{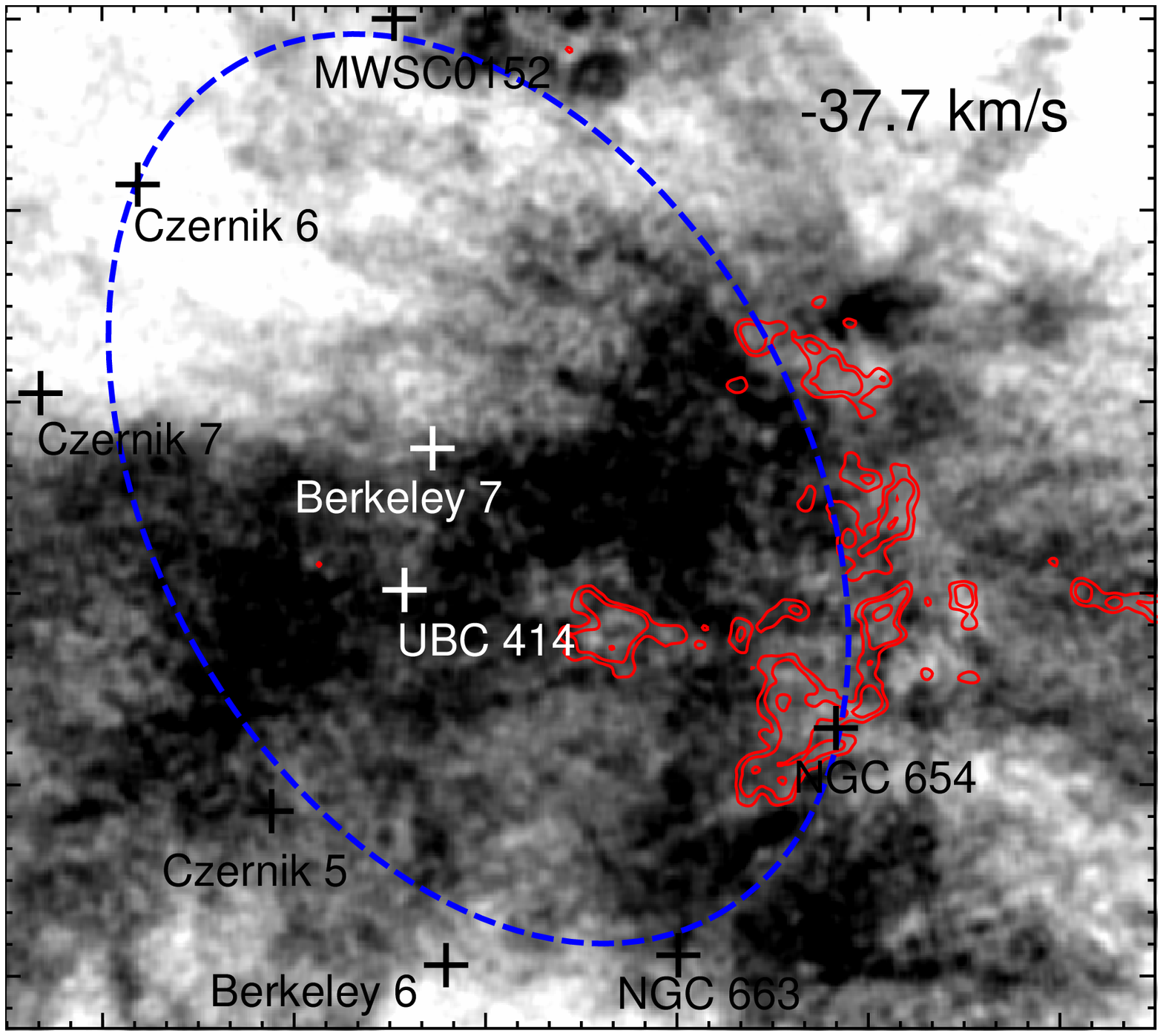}
	\includegraphics[width=4.3cm]{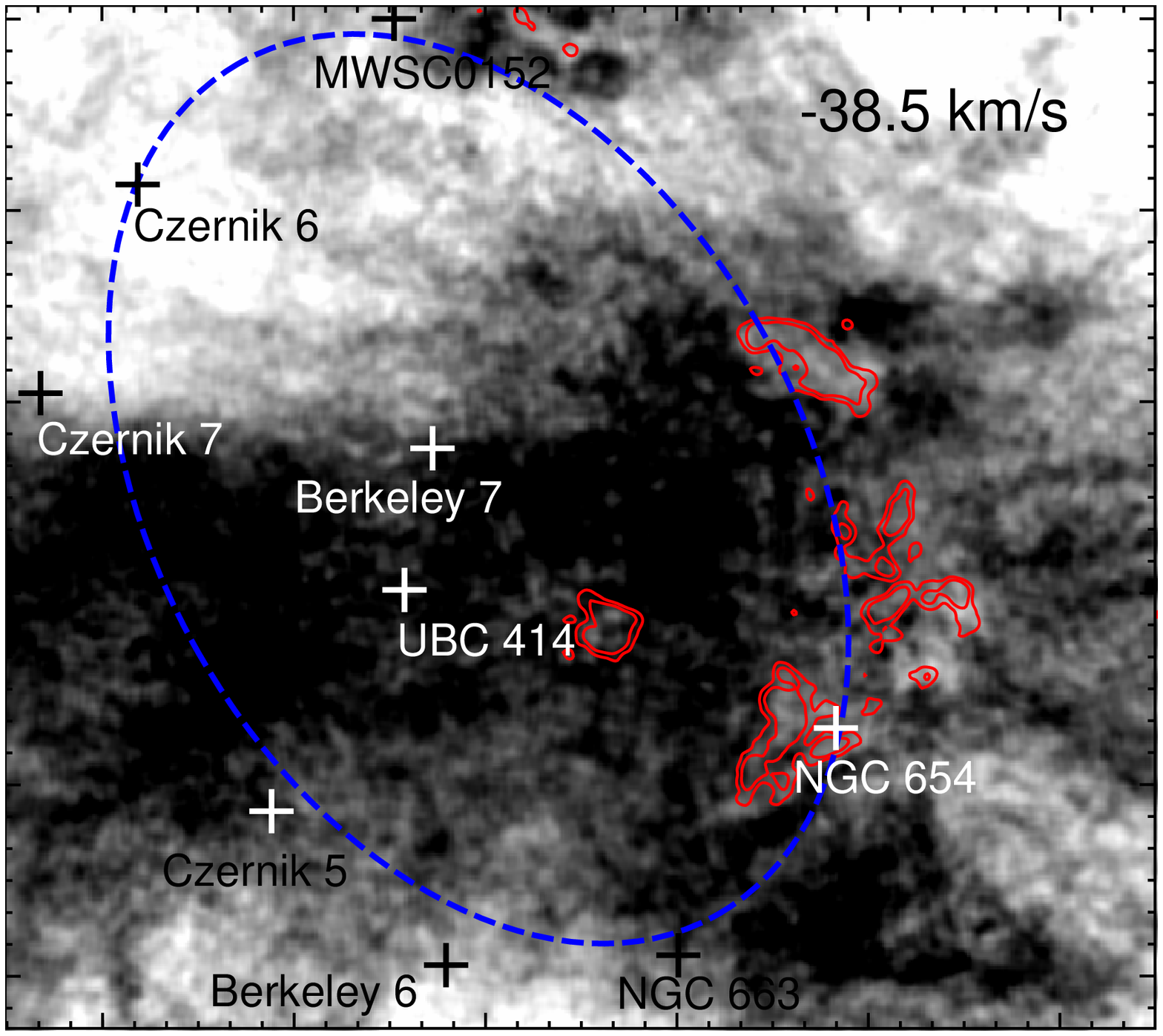}
	\includegraphics[width=4.3cm]{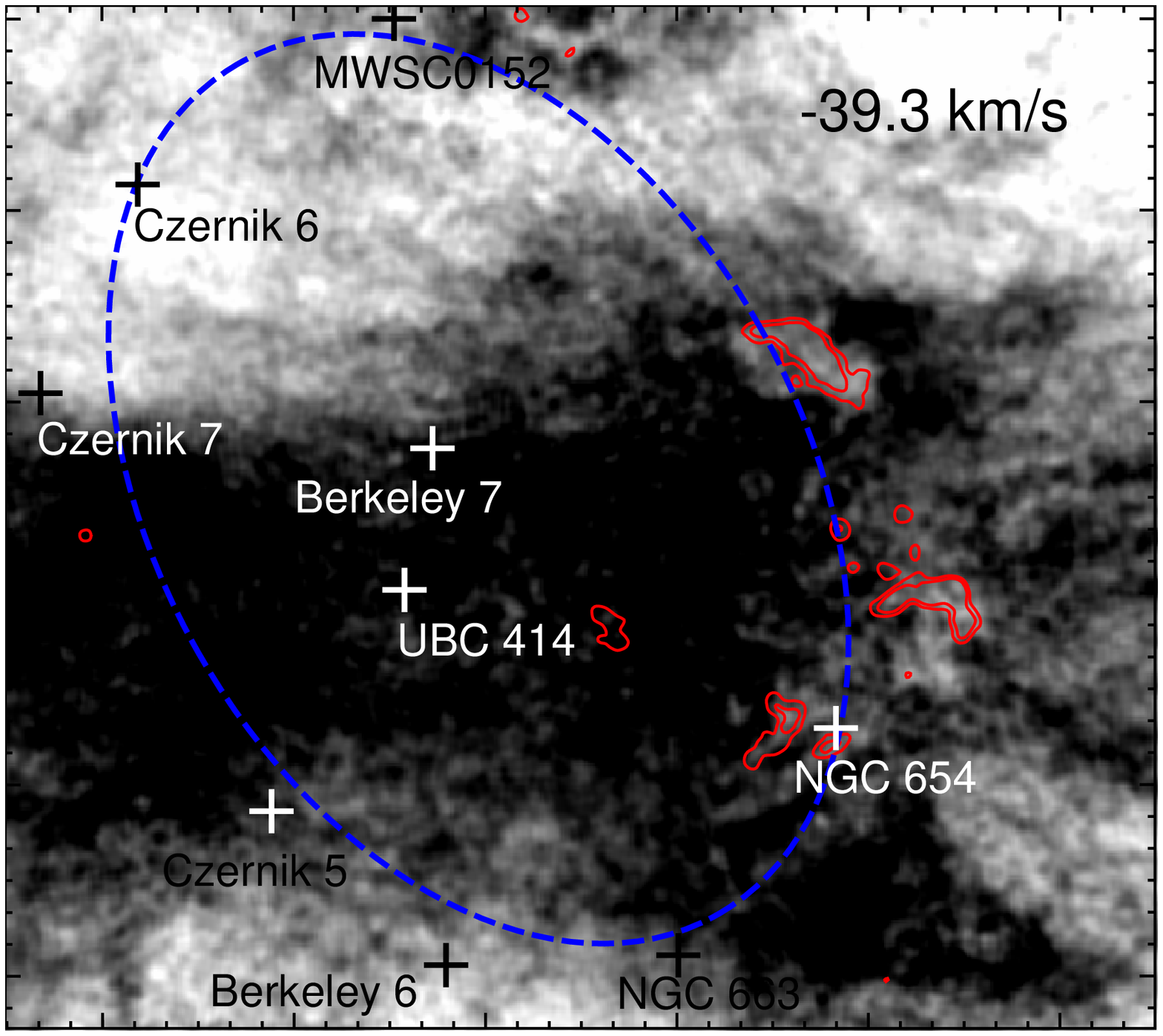}
	\includegraphics[width=4.3cm]{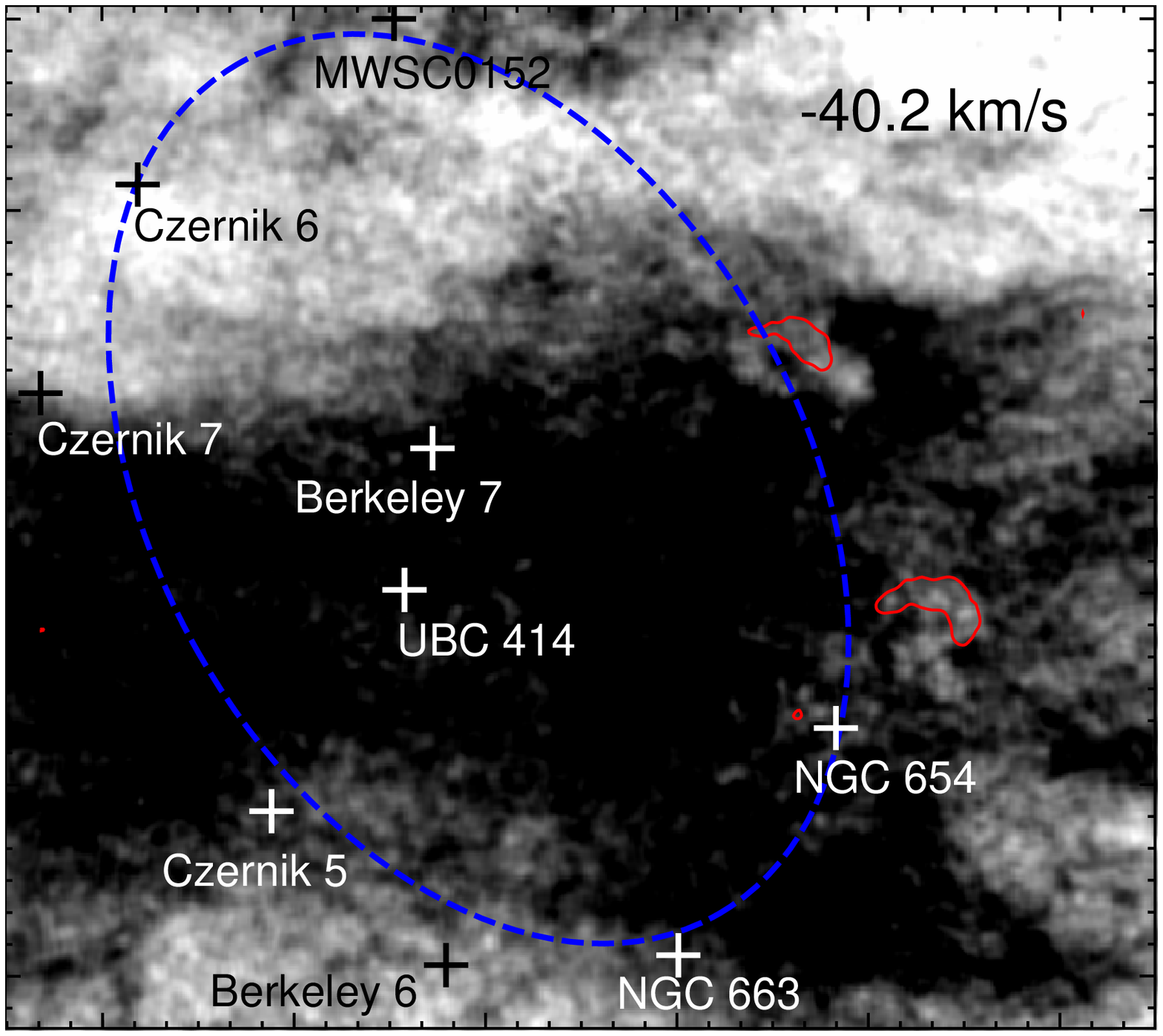}	
    \caption{Same as Fig.\,\ref{chann1} but in the velocity interval $-34.4$ to $-40.2$ \ks, and with a different range
    in the grayscale (see the colourbar presented at the first panel).}
    \label{chann2}
\end{figure*}

Figures\,\ref{chann1} and\,\ref{chann2}~present the \hi~and \2 emission displayed in channel maps going from $-27.8$ to $-40.2$ \ks, where the 
distribution of the atomic shell and the fragmented molecular cloud along the velocity interval can be appreciated. In the case
of the molecular gas emission, we can see that it appears clumpy not only in the plane of the sky but also along the line of sight. The molecular fragments appear over the \hi~shell suggesting an old expansion event that fragmented a molecular cloud.
Some of the main molecular clumps and fragments seem to be located at the farthest side of the atomic shell along the line of sight.
Given the Galactic position and the distances (about 3 kpc), it is important to remark that from $\sim -37$ \ks~to larger velocities, the atomic emission should correspond to the emission from the Perseus Galactic Arm (see the bulk and saturated emission displayed mainly at the last four maps). Nevertheless, the nature of an atomic shell is evident along the velocity interval. Despite the confusion with
gas that may correspond to the mentioned Galactic arm, it can be appreciated that the size of the empty inner region of the shell decreases towards larger velocities, i.e. towards regions representing the farther border of the shell along the line of sight, as it can be expected in this kind of bubbles/shells.

In the following section we perform an analysis of the kinetic energy stored in the shell.

\subsection{Kinetic energy stored in the \hi~shell}
\label{energy}

Firstly we calculate 
the \hi~column density from the typical equation: ${\rm N(\hi)} = 1.823 \times 10^{18} \int^{v_2}_{v_1} {\rm T_{B}}~dv $ cm$^{-2}$, where
$v_1$ and $v_2$ are $-28$ and $-36$ \ks, the velocity range along with the shell extends, and T$_{\rm B}$ is the brightness temperature. Then, the gaseous mass associated with the \hi~shell was obtained from a similar expression as used in the estimation of the molecular mass (Sect.\ref{cocloud}).
In this case, the summation was done over all the area delimited by 280 K \ks~contour (see Fig.\ref{figHI}). Assuming a distance of $2.8$ kpc and a helium
abundance of 30\%~by mass, the total gaseous mass of the \hi~shell is about $3.2 \times 10^5$ \msol. 

Considering that the expansion velocity of the shell is $v_{\rm exp} = 4$ \ks~($v_{\rm exp} = \Delta v/2$,
where $\Delta v$ is the velocity range along which the structure is observed), the kinetic energy
stored in the shell can be estimated from: $E_{\rm kin} = 0.5~{\rm M}~v_{\rm exp}^2$, yielding an
energy of 5 $\times 10^{49}$ erg. If the mass of the molecular clouds described in Sect.\,\ref{cocloud} is also considered, the kinetic energy stored in the shell is about $8.3 \times 10^{49}$ erg. Regarding the involved errors, considering a conservative 
error in the distance of 10\%, an error 
of about 20\% is derived for the kinetic energy stored in the shell, which yields an upper value of $E_{\rm kin} = 1 \times 10^{50}$ erg. The mass value and
the $E_{\rm kin}$ are consistent with others \hi~shells studied in the literature (e.g. \citealt{mcclure02,suad16}). 

Additionally, defining the dynamic age of the structure as $t_{\rm dyn} = R_{\rm eff}/v_{\rm exp}$, which is a rough estimate of the time along which the energy was injected in the shell to form it, where $R_{\rm eff} = 47.5$ pc is the effective radius of the shell ($R_{\rm eff} = \sqrt{a b}$, where $a$ and $b$ are the semi-axes of the ellipse), we obtain $t_{\rm dyn}$ about 11 Myr.

\subsection{The origin of the shell}

In this section, we analyse whether the origin of the shell
can be attributed to the action of stellar winds of massive stars 
located within the shell, eventually belonging to Berkeley~7 and/or UBC 414 clusters.

Firstly, we notice that the galactic Cartesian coordinates of both clusters \citep{can20}, indicate that their centers are separated only 69 pc, which is within the typical spatial linking length of open cluster pairs \citep[100 pc,][]{Conrad2017}. Such small physical separation, together with their similar mean proper motions, could indicate a possible common origin or physical interaction between Berkeley 7 and UBC 414, and certainly deserves further studies, but is beyond the scope of the present article.

The open cluster Berkeley 7 lies at the center of the shell, and according to \citet{can20} it has an age of 34~Myr.  
The authors list 89 stars as possible members of Berkeley~7, which according to their GAIA magnitudes, their age, their distance, and considering E(B-V)=0.7 \citep{Molina2018} are in the range 1.3 M$_\odot$ to around 9 M$_\odot$, following GENEVA rotating isochrones \citep{eks12}, reddened considering the laws given by \citet{Babusiaux2018}. The two brightest stars of the cluster are LS I +62 191 and LS I +62 190, classified as B1.5III and B2V respectively, by \citet{Molina2018}. The third brightest star, TYC 4036-1237-1, has not been assigned a spectral type yet, but its position in the colour-magnitude diagram, very close to LS I +62 190, points to the same spectral type, B2V.

UBC 414 is a recently detected open cluster \citep{Castro-Guinard2020}, gathering 29 confirmed stellar members, with age and distance similar to those of Berkeley 7: 30 Myr and 2.9 kpc. Considering these cluster parameters and Geneva isochrones \citep{eks12}, reddened considering the laws given by \citet{Babusiaux2018},  we find E(B-V)=0.55.  According to this isochrone, only the four brightest objects have masses larger than 7 M$_\odot$, and could blow stellar winds. These stars are 
TYC 4036-1373-1 (G=10.94), EM$^*$ GGA 128 (G=11.56), TYC 4036-999-1 (G=11.98), TYC 4036-2029-1 (G=12.09). Among them, only EM$^*$ GGA 128 has been studied in the past, though not thoroughly, and was assigned a spectral type OBe by \citet{Hardorp1959}.  Being similar in distance and age, the 4 brightest stars in UBC 414 seem to be of similar spectral type as those of Berkeley 7, around B2V or earlier.

Therefore, we can consider that at least these seven B2V stars could have contributed to the generation of the
\hi~shell. Following \citet{leith92} and \citet{lamers93}, we use a mass loss rate and a wind velocity of $\dot{M} = 10^{-8.8}$ \msol~yr$^{-1}$ and $v_{w} =$ 2200 \ks~, typical values for a 10 \msol~star, 
to calculate its wind energy from $E_{w} = 0.5~\dot{M}~v_{w}^2~\tau$. Assuming $\tau$ about 30 Myr, the age of the clusters, we obtain $E_{w} = 1.6 \times 10^{49}$ erg injected by the seven 10 \msol~stars, which is clearly not enough to explain the kinetic energy stored in the \hi~shell (see Sect.\,\ref{energy}). Moreover, using the dynamic age of the \hi~shell as $\tau$, $E_{w}$ is even lower. However it is pretty plausible that other undetected massive stars lie, or have lied, in the region. For instance, three O7V or ten O9V stars would produce a wind energy of about $4.2 \times 10^{50}$ erg, and considering a theoretical energy conversion ($\epsilon=E_{k}/E_{w}$) of 0.2 \citep{weav77}, they could explain the kinetic energy stored in the shell. 

By assuming an initial mass function (IMF), we can roughly estimate the population of massive stars that clusters Berkeley 7 and UBC 414 could have harbored. We consider a typical IMF with N(M) $= N_0$ M$^{-2.35}$ \citep{Salpeter1955,Kroupa2001}, the ages and distances of Berkeley 7 and UBC 414, and isochrones of \citet{eks12}, and we assume that the present day observed number of stars (that correspond to the mass range 1.21 -- 9.1 \msol) is representative of that of the IMF. Under these assumptions, the number of stars with masses larger than 9.1 \msol~at the ZAMS for Berkeley 7 is 6, and for UBC 414 is 2.

Then, we estimate that besides the seven massive stars described above, the clusters Berkeley 7 and UBC 414 together could have had about 
eight stars with M $>$ 9 \msol~in the past. These stars could have contributed to the generation of the \hi~shell, and moreover, some of them could have exploded as a supernova and injected mechanical energy through their shock fronts (e.g. \citealt{norma00,rel07,henn14}). Additionally, we cannot discard that
this shell can be a remnant of a past cloud-cloud collision event, which, as several authors propose, may be 
efficient in star formation (\citealt{fukui21} and references therein).

In the following section, we explore whether the expansion of the shell could affect the stellar population of the clusters located at its borders.

\section{Relating the shell and the clusters}

The two youngest clusters located over the shell-like structure are MWSC0152 and NGC~654 with about 4 and 10~Myr old, respectively. Taking into account the shell dynamical time of about 11 Myr, we suggest that the
formation of the stars belonging to these clusters could be triggered by the expansion of this shell \citep{elme02,elme12}. Moreover, in the case of NGC~654, it is interesting to note that the cluster is related to a molecular clump, showing that the parental molecular cloud has not been completely dispersed yet.

The oldest cluster of the group is Czernik 7 (about 1000 Myr). Given to its age and that it is probably located at a farther distance, it is unlikely a dynamical association with the shell.
Another old cluster, but in this case very likely related to the shell, is Berkeley 6, with 220~Myr. Based on its age and the low amount of stellar members (79 stars, according to \citealt{can18}) it is unlikely that star formation induced by the shell expansion has occurred within it. 

Finally, Czernik 6 and NGC 663 appear to have ages similar to those of Berkeley 7 and UBC 414, around 30 Myr, and it is possible that they have been affected by the expansion of the shell. 
While Czernik 6 hosts only a few tens of stars, the open cluster NGC 663 has more than 620 members \citep{can18}. Then, this cluster is massive enough to explore a possible scenario of triggered star formation favoured by the expansion of the shell due to the action of
Berkeley7/UBC414. Certainly, NGC 663 lies in a region very bright at 12 $\mu$m (see Fig.\,\ref{figir12}), showing the presence of abundant dust which supports a scenario of star formation. 

In order to study possible different stellar populations within NGC 663, we investigate
the kinematical properties of the stellar members using the new GAIA EDR3 data, which provides improved proper motions. 
By considering as cluster members of NGC 663 those objects with cluster membership probability larger than 0.5 according to \citet{can18}, we obtain GAIA EDR3 mean values for the cluster position RA. 26\fdg57079, dec. 61\fdg21464, a cluster distance of 2912 pc (corresponding to the median of parallax values of 0.34345) or 2926 pc (mean value of individual stellar distances), and proper motion $\mu_{\rm RA}^{*}=-1.13649554$ mas yr$^{-1}$ and $\mu_{\rm dec}=-0.33283439$ mas yr$^{-1}$. While the position and distance of the stars remain unchanged between GAIA DR2 and EDR3 data releases, the proper motions have improved.

\begin{figure}
    \centering
	\includegraphics[width=8.3cm]{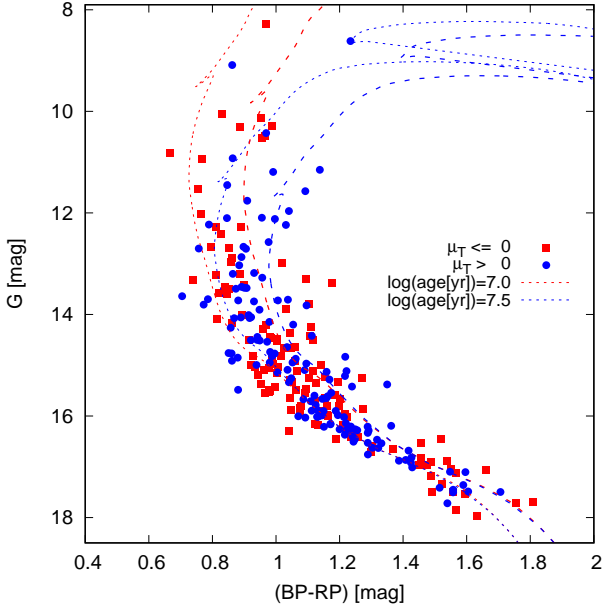}
    \caption{GAIA DR2 CMD of the central part of NGC 663, the most densely populated region where half of the stars are located. We plot stars with $\mu_T$<0 (red) and $\mu_T$>0 (blue). Blue curves indicate Geneva isochrones for rotating populations \citep{eks12}, corresponding to the age of the cluster quoted in the literature (30 Myr, see Table 1), with two different colour excesses: 0.75 (dotted line) and 0.85 (dashed line), typical intra-cluster excesses  given in the literature for NGC 663 \citep[e.g.][]{Phelps1994}. Interestingly, stars with $\mu_T$<0 seem to be younger than those with $\mu_T$>0, with an isochrone age around 10 Myr (red curves).}
    \label{cmd}
\end{figure}

We obtain cartesian postions and proper motions relative to the mean values described above \citep[following e.g.][]{Helmi20}, which allows us to derive accurate tangential and radial components of the proper motion in the plane of the sky ($\mu_T$ and $\mu_R$) for each star. In particular, $\mu_T$ results of interest because it is indicative of intra cluster rotational motions. 
Then, we explore the GAIA DR2 color-magnitude diagram of the central part of the cluster, the most densely populated region where half of the members and most massive stars are found and where the emission at IR is more intense, and therefore, star formation is most likely to occur. We consider rotating isochrones \citep{eks12}, reddened taking into account the extinction laws for GAIA DR2 data \citep{Babusiaux2018}, and we interestingly find that stars with $\mu_T$>0 correspond to a population that is globally older than that with $\mu_T$<0 (see Fig.\,\ref{cmd}). Therefore, we propose
that NGC 663 has two different stellar populations regarding their ages, in agreement with works that point out that the
simple stellar populations scenario does not seem to hold
for some intermediate-age and young star clusters \citep{valcarce11,li16}. It is important to remark that the age of the youngest population (10 Myr) coincides with the dynamical time of the shell, strongly suggesting 
a scenario in which the already existing cluster NGC 663, was reached by the expanding shell, triggering the formation of these stars, that share a similar motion. A detailed study of NGC 663 and its interesting populations is deferred to a future study.

\section{Summary and concluding remarks}

We investigated a Galactic region ($l=130\fdg0, b=0\fdg35$) populated by several
open stellar clusters. We noticed that most of the clusters in the region (MWSC0152, Czernik 6, Berkeley 6, NGC 663, NGC 654, Berkeley 7, and UBC 414) are located almost at the same distance (about 2.9 kpc), and also the distance of Czernik 7 can be considered close to 3 kpc. Given that clusters Berkeley 7 and UBC 414 appear at a position around which the other clusters are distributed quite equidistantly, we studied the ISM in order to search for some evidence pointing to a connection among them.

By analyzing the ISM at IR, centimeter, and millimeter wavelengths we discovered a shell 
of interstellar material in whose borders the clusters MWSC0152, Czernik 6, Czernik 7, Berkeley 6, NGC 663, and NGC 654 lie. From the
mid-IR emission we found some interesting features such as pillars and globules strongly suggesting an interaction between
the radiation from massive stars and molecular gas. From the kinematical analysis of the \hi~and CO emissions we found that the shell is located at the same distance as the clusters, and it has a mass of about $3\times 10^{5}$ \msol. We estimated that the kinetic energy stored in the shell is in the range $(0.8-1.0)\times 10^{50}$ erg, and its dynamical age is about 11 Myr.  We point out that it is very likely that massive stars belonging 
to Berkeley 7 and UBC 414 and/or supernova events occurred in these clusters have generated this shell and influenced the stellar population of the clusters lying at its borders. 

By comparing the ages of the shell and the clusters, we suggest that the expansion of the shell could have triggered
the formation of the stars of  MWSC0152 and NGC~654. In the case of the latter, the youngest among all the clusters in the region, it is important to note that it is still related to a molecular cloud, supporting a triggered star formation scenario in which the parental
molecular cloud has not been dispersed yet.
Cluster NGC 663 was of particular interest. We discovered that this cluster contains two different stellar populations regarding their ages: a group of stars with about 30 Myr old, and another group with about 10 Myr, strongly suggesting that the expansion of the shell triggered the formation of this second group.

In this work we presented solid evidences pointing to probe triggered star formation among stellar clusters at large spatial scales, and it motivates further studies to other stellar clusters groups in order to find a genetical connection among them.

\section*{Acknowledgements}

We aknowledge the anonymous referee for her/his useful comments. M.B.A. is a doctoral fellow of CONICET, Argentina.
S.P. and A.G. are members of the {\sl Carrera del Investigador Científico} of CONICET, Argentina.
A.G. acknowledges the financial support received from the Agencia Nacional de Promoción Cientifíca y Tecnológica of Argentina (PICT2017-3790).
This work has made use of data from the European Space Agency (ESA) mission
{\it Gaia} (\url{https://www.cosmos.esa.int/gaia}), processed by the {\it Gaia}
Data Processing and Analysis Consortium (DPAC,
\url{https://www.cosmos.esa.int/web/gaia/dpac/consortium}). Funding for the DPAC
has been provided by national institutions, in particular the institutions
participating in the {\it Gaia} Multilateral Agreement.

\section*{Data availability}

The data underlying this article were accessed from: \\ 
https://irsa.ipac.caltech.edu/Missions/wise.html,\\ 
https://www.cadc-ccda.hia-iha.nrc-cnrc.gc.ca/en/search/, and \\
https://www.cosmos.esa.int/web/gaia/data.



\bibliographystyle{mnras}
\bibliography{ref.bib} 








\bsp	
\label{lastpage}
\end{document}